# Polarization-Selective Control of Nonlinear Optomechanical Interactions in Subwavelength Elliptical Waveguides

Dae Seok Han,[1] Il-Min Lee,[2] Kyung Hyun Park,[2] and Myeong Soo Kang[1,*]

[1]Department of Physics, Korea Advanced Institute of Science and Technology, 291 Daehak-ro, Yuseong-gu, Daejeon 34141, Republic of Korea
[2]Terahertz Basic Research Section, Electronics and Telecommunications Research Institute, 218 Gajeong-ro, Yuseong-gu, Daejeon 34129, Republic of Korea
*Corresponding author: mskang@kaist.ac.kr

**Photonic devices exhibiting all-optically reconfigurable polarization dependence with a large dynamic range would be highly attractive for active polarization control. Here, we report that strongly polarization-selective nonlinear optomechanical interactions emerge in subwavelength waveguides. By using full-vectorial finite element analysis, we find that at certain core ellipticities (aspect ratios) the forward simulated light scattering mediated by a specific acoustic mode is eliminated for one polarization mode, whereas that for the other polarization mode is rather enhanced. This intriguing phenomenon can be explained by the interplay between the electrostrictive force and radiation pressure and turns out to be tailorable by choice of waveguide materials.**

Photonic devices that exhibit polarization dependence are detrimental in general, as they degrade the performance of photonic systems when employed. On the other hand, if the polarization dependence could be all-optically controlled with a large dynamic range at modest optical power levels, it would be highly attractive for active control and manipulation of the polarization state of light, which can be adopted in new types of photonic components and advanced all-optical signal processing. Although stimulated Brillouin scattering (SBS) in single-mode fibers has been considered as one of the most significant nonlinear optical effects [1,2], the insufficient polarization dependence of SBS gain [3,4] has still hindered its practical use for implementing all-optical polarization manipulation. Strongly polarization-selective SBS amplification of one particular polarization mode could also dramatically improve the performances of various SBS-based systems, e.g. tunable bandpass filters [5], Brillouin-based optical spectrometers [6], and optical vector network analyzers [7].

In micro/nano-scaled waveguides, on the other hand, light and acoustic phonons can be confined simultaneously in tiny spaces. Novel kinds of nonlinear optomechanical phenomena can then emerge, as recently demonstrated in micron/submicron-thick fiber tapers [8–10], small-solid-core microstructured fibers [11–13], and silicon on-chip suspended waveguides [14–17]. In these systems, tailored dispersions of trapped acoustic phonons give rise to forward SBS (FSBS) via the phase-matched nonlinear coupling between the co-propagating guided light and acoustic resonances (ARs) [12,13]. Furthermore, when the guided light strongly interacts with the waveguide boundaries, the radiation pressure is radically enhanced, which can contribute significantly to FSBS [17]. Although tailoring of acoustic phonons and optical forces could be exploited to engineer the polarization dependence of photon-phonon interactions, previous works have been limited mostly to changing the cross-sectional waveguide dimensions to adjust the Brillouin frequency shift and gain [8–10,14–16]. It might be anticipated that the photon-phonon interactions can be highly polarization-sensitive in subwavelength-scaled photonic systems with a strong lack of the circular or C4 symmetry, although such the possibility has not been studied yet.

In this Letter, we show that the photon-phonon interactions, particularly FSBS, can be strongly 'polarization-selective' in subwavelength waveguides with suitably designed core geometry, which provides a novel way of highly efficient all-optical dynamic polarization control. By carefully selecting core ellipticities (or aspect ratios) and dimensions, we can completely suppress the FSBS mediated by a certain acoustic phonon mode for only one optical polarization mode, while keeping that significant for the other. We perform full-vectorial simulations of optical and acoustic modes in the waveguides and investigate their nonlinear interactions to fully understand the polarization selectiveness.

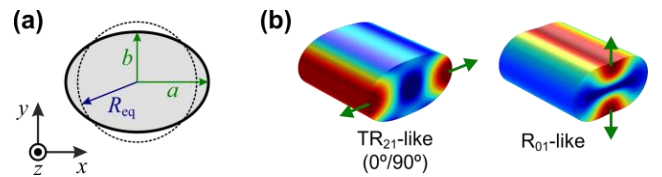

Fig. 1. (a) Cross-section of a typical elliptical waveguide suspended in the air, together with the Cartesian coordinates. The equivalent radius $R_{eq}$ represents the core dimension, in such a way that the cross-sectional area of the core (grey ellipse) equals that of the circular core of radius $R_{eq}$ (dotted circle). (b) ARs in elliptical waveguides that mediate FSBS significantly. A color map is used to describe the profile of magnitude of total displacement, where blue and red correspond to the zero and maximum displacement, respectively. Green arrows indicate the direction of acoustic displacement.

We consider silica-glass elliptical waveguides suspended in the air, as described in Fig. 1(a). This waveguide structure is practically feasible in the form of highly birefringent microstructured fibers [18], elliptical microfibers [19], and air-suspended on-chip slab waveguides [14–16]. The photon-phonon interactions in the elliptical waveguide depend generally on both the ellipticity and dimension of the core. Here, we define the core ellipticity as $e = (a–b)/a$ ($0 \leq e \leq 1$), where $a$ and $b$ are the semi-major and semi-minor axis, respectively. We use the 'equivalent radius' $R_{eq} = \sqrt{ab}$ to represent the core dimension. By considering the symmetry of the optical modes, it can be verified that the $TR_{21}$-like (0°/90°) torsional-radial AR and the $R_{01}$-like radial AR (Fig. 1(b)) can drive intramodal FSBS significantly [12,13]. We note that rectangular waveguides could be an alternative model of realistic waveguides having the C2 symmetry, while the polarization sensitivity of FSBS has not been investigated in the previous experiments with on-chip rectangular waveguides [14-16]. Our analysis on the rectangular waveguides, not to be presented here, shows that their key qualitative features of polarization selectiveness are almost the same as those of elliptical ones.

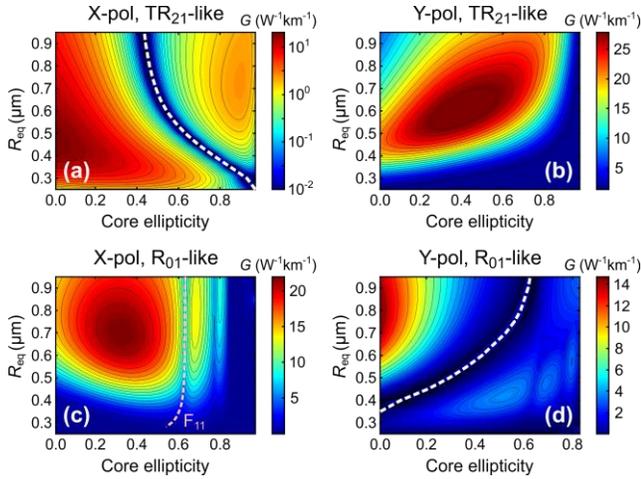

Fig. 2. Contour plots of the FSBS coupling in silica glass elliptical waveguides as functions of the equivalent radius ($R_{eq}$) and ellipticity of the core for each AR and polarization mode. The white dashed curves in (a) and (d) indicate the core parameters at which the FSBS coupling is eliminated. The pink dashed curve in (c) follows the acoustic anti-crossing points between the $R_{01}$-like and the flexural $F_{11}$ ARs. Notice that the plot in (a) is in the logarithmic scale.

We define 'FSBS coupling', a figure of merit for FSBS, as $G = g_0/Q_m$, where $g_0$ and $Q_m$ are the FSBS gain and the quality factor of the AR, respectively. We calculate the intramodal FSBS coupling [13,20] at the wavelength of 1550 nm for the two types of ARs, over the entire range of ellipticity and equivalent radius of the core (Fig. 2). The photon-phonon interactions via intramodal FSBS can be highly polarization-selective, in the sense that the FSBS coupling is eliminated for one polarization mode, while it is kept significant for the other. The strong polarization selectiveness is observed over a wide range of core dimension for both ARs. For instance, for the $TR_{21}$-like AR, the FSBS driven by the x-polarized mode is completely suppressed at certain core ellipticities, which appears as a zero-FSBS-coupling curve in Fig. 2(a). In sharp contrast, around the zero-FSBS-coupling conditions, the FSBS coupling for the y-polarized mode is rather enhanced, having a maximum value of $G = 29$ W$^{-1}$km$^{-1}$ at ($R_{eq}$, $e$) = (607 nm, 0.42), as can be seen in Fig. 2(b). For the $R_{01}$-like AR, on the contrary, the FSBS coupling is eliminated only when driven by the y-polarized mode, as shown in Figs. 2(c) and 2(d).

To figure out the strong suppression of FSBS, we take into account the contribution of electrostriction and that of radiation pressure separately to the total FSBS coupling. We define $G^{(es)}$ and $G^{(rp)}$ as the FSBS coupling components independently contributed by the electrostrictive force and radiation pressure, respectively, and calculate them at $R_{eq}$ = 500 nm over the entire range of core ellipticity. At the core ellipticity for which FSBS is suppressed, the two FSBS coupling components have the same magnitude as each other as shown in Figs. 3(a) and 3(d), which implies that they are cancelled out to yield the zero total FSBS coupling.

The strong polarization selectiveness of FSBS can be intuitively understood in terms of the time-averaged optomechanical work density $W \propto \text{Re}(\mathbf{f}^* \cdot \mathbf{u}_m)$ done on the waveguide by the optical forces, as described in Figs. 3(e-n), where $\mathbf{f}$ and $\mathbf{u}_m$ are the optical force distribution and the displacement profile of AR, respectively. The optical forces are almost transverse in the FSBS process, and the dominant electrostriction stress tensor components are then $\sigma_{xx}$ and $\sigma_{yy}$ for both polarization modes, which are expressed for the electric field profile $\mathbf{E}$ in isotropic media by [21]

$$\sigma_{xx} = -\tfrac{1}{2}\varepsilon_0 \cdot n^4 \left[ p_{11}|E_x|^2 + p_{12}\left(|E_y|^2 + |E_z|^2\right) \right], \quad (1)$$

$$\sigma_{yy} = -\tfrac{1}{2}\varepsilon_0 \cdot n^4 \left[ p_{11}|E_y|^2 + p_{12}\left(|E_x|^2 + |E_z|^2\right) \right], \quad (2)$$

where $\varepsilon_0$ and $n$ are the vacuum electric permittivity and the refractive index, respectively, and $p_{11}$ and $p_{12}$ are the photoelastic coefficients (PECs). For fused silica glass ($p_{11}$ = 0.121 and $p_{12}$ = 0.270), both PECs are positive and $p_{11} < p_{12}$ then makes $\sigma_{yy}[\sigma_{xx}]$ dominant over $\sigma_{xx}[\sigma_{yy}]$ for the x[y]-polarized mode. The electrostrictive bulk force $F_i^{(es,bulk)} = -\partial \sigma_{ij}/\partial x_j$ and boundary force $F_i^{(es,boundary)} = \sigma_{ij}n_j$ [17] are then applied inward, where $n_j$ is the normal vector component at the waveguide surface. In addition, the electrostrictive bulk force is mostly perpendicular to the optical polarization. While for the x-polarized mode the electrostrictive bulk and boundary forces exerted together in the y-direction get significantly greater than the counterbalancing radiation pressure (Figs. 3(g,i)), for the y-polarized mode the radiation pressure is dominant (Figs. 3(h,j)). As a result, the total optical force distribution created by the y-polarized mode exhibits a 'squeezing' pattern over the waveguide cross-section that closely resembles the $TR_{21}$- like AR, which yields a significant work (Fig. 3(l)). The y-polarized mode is then coupled efficiently to the TR$_{21}$-like AR, resulting in the enhancement of FSBS coupling. On the contrary, when driven by the x-polarized mode, the total optical force distribution does not match the TR$_{21}$-like AR (Figs. 3(k,m)), and the resulting work can then be cancelled out, which gives rise to the suppression of FSBS coupling. The opposite behavior of polarization-selective FSBS suppression for the R$_{01}$-like AR can be explained in a similar fashion.

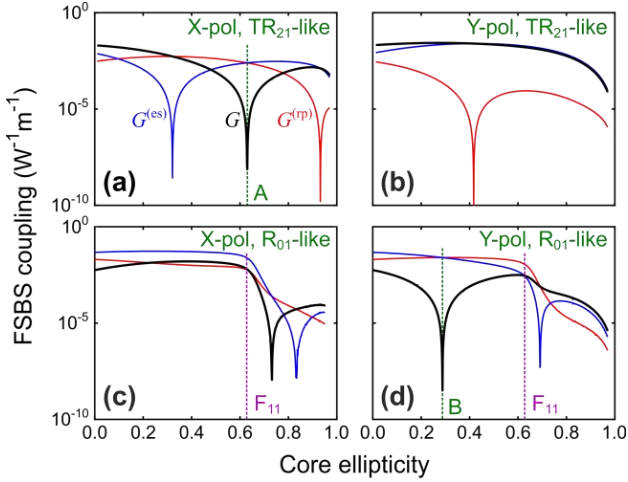

Fig. 3. Explanation of the polarization-selective FSBS in silica glass elliptical waveguides. (a-d) Contribution of electrostriction (blue curves) and radiation pressure (red curves) to the total FSBS coupling (black curves) at $R_{eq}$ = 500 nm over a range of core ellipticity for each AR and polarization mode, which correspond respectively to Figs. 2(a-d). The green vertical dashed sections indicated by 'A' in (a) and 'B' in (d) point to FSBS suppression. At the purple vertical dashed sections in (c) and (d), acoustic anti-crossing emerges between the R$_{01}$-like and the flexural F$_{11}$ ARs. (e-n) The optical field profiles, optical force distributions, and the resulting optomechanical work densities on the waveguide cross-section, in the condition where FSBS mediated by the TR$_{21}$-like AR is suppressed for the x-polarized mode ($e$ = 0.63). The electrostrictive force ($\mathbf{f}^{(es)}$) and the radiation pressure ($\mathbf{f}^{(rp)}$) are shown in blue and red, respectively. The bulk and boundary forces are plotted in the same scale for comparison. The positive and negative work densities are displayed as red and blue, respectively.

We also obtain the spectra of FSBS for each polarization mode by calculating the acoustic frequency and FSBS coupling for several ARs (including the TR$_{21}$-like and the R$_{01}$-like ones) over the entire range of core ellipticity, while keeping the core dimension fixed ($R_{eq}$ = 500 nm for Fig. 4). It is noteworthy that a number of anti-crossings emerge at some core ellipticities, which we attribute to the simultaneous resonances of two acoustic modes satisfying the free-boundary (Neumann) conditions for the acoustic displacement at the waveguide interface [22]. For instance, the frequency of the R$_{01}$-like AR increases with the core ellipticity and intersects that of the higher-order flexural F$_{11}$ AR at $e$ = 0.63. The amount of frequency splitting at the resulting anti-crossing depends on the coupling strength between the two ARs. As the core ellipticity increases further, the R$_{01}$-like AR branch forms a series of anti-crossings with other types of ARs. We collect these R$_{01}$-like ARs and designate them here as a 'family of R$_{01}$-like ARs'. We note that for the x-polarized mode the FSBS coupling decreases significantly nearby the acoustic anti-crossings, as can be seen in Fig. 2(c). In addition, for the y-polarized mode, the flexural ARs dominate the FSBS coupling over the R$_{01}$-like ones at the core ellipticities above the first anti-crossing, which results in negligible FSBS couplings of the R$_{01}$-like ARs at high core ellipticities (Fig. 2(d)).

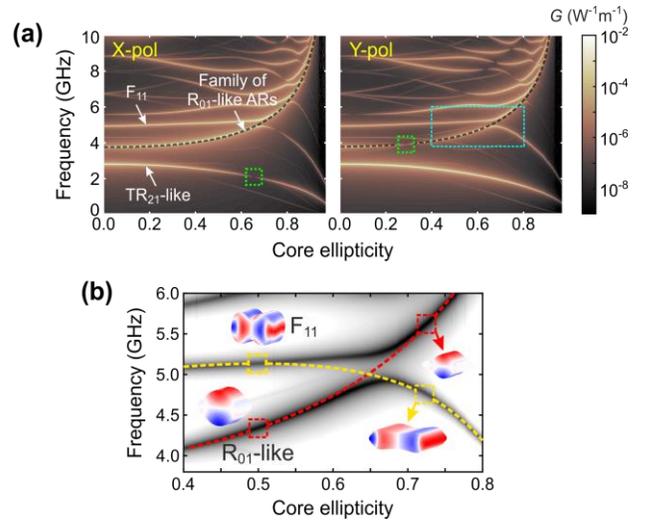

Fig. 4. (a) FSBS spectra of silica glass elliptical waveguides of $R_{eq}$ = 500 nm over a range of core ellipticity, for each polarization mode. The black dashed curves indicate a family of R$_{01}$-like ARs in virtue of the strong coupling

between torsional and radial displacements. The green dashed squares show FSBS suppression. (b) Zoomed-in FSBS spectra corresponding to the light blue dashed rectangle in (a). The white and black in the color map correspond to the zero and maximum value of FSBS coupling, respectively. Two dashed curves represent the $R_{01}$-like (red) and the flexural $F_{11}$ (yellow) ARs, and the inset shows their displacement profiles with the exaggerated deformation for clarity, where the color map represents the y-displacement ($u_y$) for the $R_{01}$-like AR and the x-displacement ($u_x$) for the $F_{11}$ AR, blue, white and red corresponding to negative, zero and positive values, respectively.

It is worth comparing the results so far for silica waveguides with those for silicon counterparts, as the latter has recently attracted rapidly growing attention, being anticipated to exhibit ultrahigh optomechanical interaction efficiencies [17]. Silicon has the PECs of $p_{11} = -0.09$ and $p_{12} = 0.017$ under the [100] orientation. On the contrary to fused silica, since $|p_{11}| > |p_{12}|$ for silicon, the electrostriction stress tensor component $\sigma_{xx}[\sigma_{yy}]$ is dominant over $\sigma_{yy}[\sigma_{xx}]$ for the x[y]-polarized mode (Eqs. (1) and (2)), and the direction of electrostrictive bulk force is then mostly parallel to the optical polarization (Figs. 5(g,h)). In addition, the negative value of $p_{11}$ makes the electrostrictive bulk force point outward (Figs. 5(g,h)), which combines constructively with the radiation pressure regardless of the optical polarization (Figs. 5(i,j)), though near the waveguide boundary relatively smaller electrostrictive forces exist that tend to pull inward against the radiation pressure (Figs. 5(i,j)).

The constructive combination of electrostrictive bulk force and radiation pressure for both polarization modes has eluded the observation of polarization selective FSBS in silicon waveguides. Indeed, the polarization-selective characteristics does not exist for the photon-phonon interactions by the $TR_{21}$-like AR (Figs. 5(a,b)). For the $R_{01}$-like AR, however, strong polarization selectiveness is observed at certain core ellipticities (Figs. 5(c,d)). For the y-polarized mode, the electrostrictive bulk force and radiation pressure cooperate in the y-direction, which in turn excite efficiently the $R_{01}$-like AR having the dominant strain component $S_{yy}$ (Figs. 5(h,j)). The time-averaged optomechanical works done on the waveguide bulk and boundary then combine in phase, which gives rise to non-zero FSBS couplings (Figs. 5(l,n)). On the other hand, when driven by the x-polarized mode, the bulk and boundary work densities are distributed in such a way that the net work done on the waveguide vanishes, yielding the FSBS suppression (Fig. 5(k,m)). We emphasize that the polarization-selective FSBS in silicon waveguides can be potentially a key phenomenon for implementing polarization devices in on-chip photonic integrated circuits.

In conclusion, we have shown that strongly polarization-selective nonlinear photon-phonon interactions emerge in subwavelength waveguides with carefully designed core geometry, which can be used as a novel way of highly efficient all-optical reconfigurable polarization control with a huge dynamic range. At certain core ellipticities (or aspect ratios), FSBS mediated by a specific AR mode is eliminated for one polarization mode, while that for the other polarization mode is rather enhanced. This intriguing phenomenon can be explained by the counterbalance between electrostriction and radiation pressure and turn out to be strongly affected by the PECs of waveguide materials. Our study provides a new opportunity of engineering boundary-enhanced optical forces and nonlinear photon-phonon interactions.

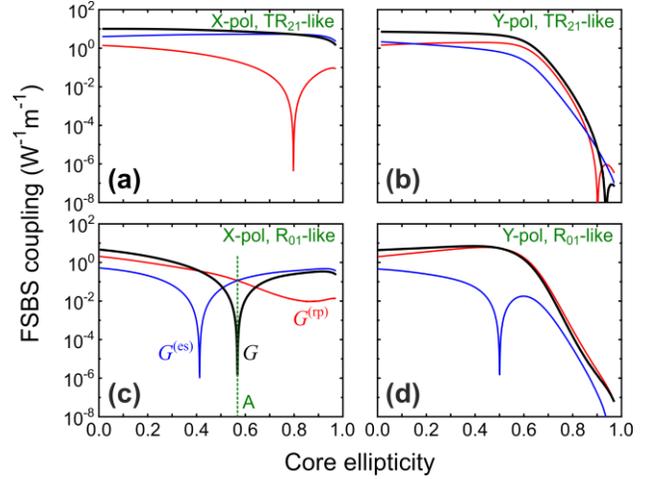

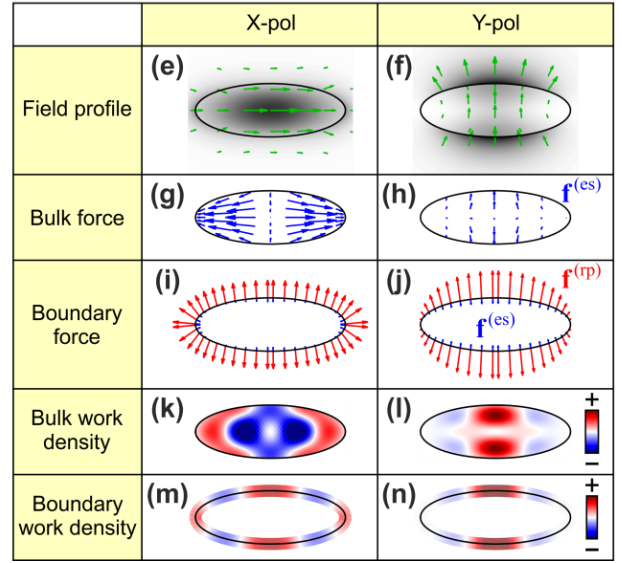

Fig. 5. (a-d) FSBS coupling in silicon elliptical waveguides as a function of the core ellipticity at $R_{eq} = 200$ nm, for each AR and polarization mode. The green vertical dashed section indicated by 'A' in (c) points to FSBS suppression. (e-n) The optical field profiles, optical force distributions, and the resulting optomechanical work densities on the waveguide cross-section, in the condition where the FSBS mediated by the $R_{01}$-like AR is suppressed for the x-polarized mode ($e = 0.55$).